\documentclass[11pt]{article}
\usepackage{fullpage,amssymb,amsmath,graphicx,natbib,hyperref}

\newcommand{\m}[1]{{\mathbf #1}}
\renewcommand{\v}[1]{\boldsymbol #1}

\newcommand{\Exp}[1]{{\rm E}[ \ensuremath{ #1 } ]  }

\newcommand{\bbeta}{\boldsymbol\beta}

\long\def\symbolfootnote[#1]#2{\begingroup
\def\thefootnote{\fnsymbol{footnote}}\footnote[#1]{#2}\endgroup}

\title{Likelihoods for fixed rank nomination networks}
\author{Peter Hoff$^{1,2}$, Bailey Fosdick$^{1}$, Alex Volfovsky$^{1}$, Katherine Stovel$^{3}$  \\ 
Departments of Statistics$^1$, Biostatistics$^2$ and Sociology$^3$ \\
University of Washington}
\date{\today}

\begin{document}

\maketitle

\symbolfootnote[0]{
This work was  supported by
NICHD grant R01 HD-67509. }

\begin{abstract}
Many studies that gather social network data use survey methods 
that lead to censored, missing or otherwise incomplete information. 
For example, the popular fixed rank nomination (FRN) scheme, 
often used in studies of schools and businesses,  asks 
study participants to nominate and rank at most a small number of 
contacts or friends, leaving the existence other relations 
uncertain. 
However, 
most statistical models  are formulated in terms 
of completely observed binary networks. 
Statistical analyses of FRN
data with such models 
ignore
the censored and ranked nature of the 
%these aspects of the survey design
%by analyzing the network data
data
% as if it were completely observed  
and could potentially result
in misleading statistical inference. 
To investigate this possibility, 
%alternative to such data analyses, 
we compare
parameter estimates obtained from a likelihood 
for complete binary networks to 
those from a likelihood that 
is derived from the FRN scheme, and therefore
recognizes the 
ranked and censored nature of the data. 
%develops a type of likelihood that is
%derived from the fixed rank nomination scheme, and therefore 
%recognizes the incompleteness present in data that are obtained  from 
%this survey design.
% We compare 
%this  fixed rank nomination likelihood to other likelihoods 
%that have been used, and 
We show analytically and 
via simulation that the binary likelihood 
can provide misleading inference, at least for certain
model parameters
that relate network ties to characteristics of 
individuals and pairs of individuals. 
We also compare these different likelihoods 
in a data analysis of several adolescent social networks.
% that exhibit 
%other types of data incompleteness. 
For some of these networks, 
the parameter estimates from the binary and FRN likelihoods  
lead to different conclusions, indicating the 
importance of analyzing FRN data with a method that accounts for 
the FRN survey design. 

\medskip

\end{abstract}

\section{Introduction}
Relating social network characteristics to individual-level  
behavior is an important application area of 
social network research. 
For example, in the context of 
adolescent health, 
many large-scale data-collection efforts have been undertaken to 
examine the relationship between adolescent 
friendship ties and individual-level behaviors, including the 
PROSPER peers study \citep{moody_2011}, 
the School Study of the Netherlands Institute for the Study of Crime and Law Enforcement (NSCR) \citep{weerman_2005}, and the 
National Longitudinal Study of Adolescent Health (AddHealth)  
study \citep{harris_2009}. 
These and other studies have reported evidence for relationships between 
friendship network ties and behaviors such as 
exercise \citep{macdonald_2011}, 
smoking and drinking behavior \citep{kiuru_2010}  and 
academic performance \citep{thomas_2000}. 

A common approach to the statistical analysis of such relationships
is via a statistical model relating the observed  social network 
data 
to a set of explanatory variables  via some unknown 
(multidimensional) parameter to be estimated. 
Often the network data are represented 
by a sociomatrix $\m S$, a square matrix 
with a missing diagonal where the $(i,j)$th element $s_{i,j}$ 
describes the relationship from node $i$ to node $j$. 
In cases where $s_{i,j}$ is the binary indicator of 
a relationship from $i$ to $j$, the sociomatrix can 
be viewed as the adjacency matrix of a directed graph.
A popular class of  models for such data are 
exponentially parameterized random graph models (ERGMs), 
typically having a small number of sufficient
statistics chosen to represent 
effects of the explanatory variables and other 
important patterns in the graph \citep{frank_strauss_1986,  
snijders_pattison_robins_handcock_2006}.
Another class of models includes 
latent variable or random effects models. These models 
assume a conditional dyadic independence in that each dyad
$\{ s_{i,j}, s_{j,i} \}$ is assumed to be independent of 
each other dyad $ \{ s_{k,l}, s_{l,k} \}$, conditional 
on some set of unobserved latent variables. The latent variables 
are often taken to be node-specific 
latent  group memberships 
\citep{nowicki_snijders_2001,airoldi_blei_fienberg_xing_2008}
or latent factors \citep{hoff_raftery_handcock_2002,hoff_2005a}, which 
can represent various patterns of clustering or dependence 
in the network data. 

While statistical models %and methods
such as these can often be very successful 
at representing the main features of a social network or relational dataset, 
they generally ignore contexts or constraints 
under which the data were gathered. 
In particular, these methods generally assume the 
relational dataset is fully observed, and that the support of the 
probability model 
is equal to the set of sociomatrices that could have been observed. 
As a simple example where such an assumption is not met, consider 
the analysis of the well-known ``Sampson's monastery'' dataset 
\citep{sampson_1969,breiger_boorman_arabie_1975}
 using 
a binary random graph model. 
These data include relations among 18 monks, each of which
was asked to nominate and rank-order three other monks whom they 
liked the most, and three other monks whom they liked the least. 
The relations reported by each monk thus consist of a partial 
rank ordering of all other monks in the monastery. 
Since a binary random graph model does not accommodate rank data, 
analysis of these 
data with such a model typically begins by reducing all 
positive ranked relations to ``ones'' and all negative 
ranked relations to ``zeros'', leaving unranked relations
as zeros as well. Such a data analysis  essentially throws
away some of  the information in the data. 
Furthermore, the support of most binary random graph models 
consists of all possible graphs on the node set. In contrast, the 
data collection scheme used by Sampson 
was censored, in that 
%implies that 
no 
graph with  more than three outgoing edges per node could have been 
observed. 

The ranked nomination scheme used to gather Sampson's monastery data 
is not exceptional. Ranked nomination methods were among the first to be used for the collection of respondent-provided sociometric data \citep{moreno_1953,moreno_1960} and have been used extensively in both research and applied settings ever since.  They remain quite common in studies of 
 work environments and 
children in classrooms,
and are recommended in 
Lawrence 
Sherman's widely used online resource ``Sociometry in the Classroom''
\citep{sherman}. 
Several large scale studies of adolescent health and behaviors 
have used variations on ranked-nomination schemes, including 
the PROSPER, NCSD and AddHealth studies mentioned above. 
However, statistical analysis of data from these studies 
generally fails to account for the ranked and censored nature of the data 
\citep{goodreau_kitts_morris_2009,weerman_2011}. % , more references needed). 

A frequent data analysis goal of many social network studies is 
to quantify  the relationships between 
ranked friendship nominations
and individual-level attributes,
such as grade level, ethnicity, academic performance and
smoking and drinking behavior. Quantification of
these relationships is 
of interest for a variety of reasons, including
identification of at-risk youth, or 
to aid in the  development of adolescent health programs,
which often have components based on peer interventions.
Statistical evaluations of the relationships are often made by 
modeling the network outcome $s_{i,j}$ for each pair of individuals $(i,j)$  as depending on a linear 
predictor $\bbeta^T\v x_{i,j}$, where 
$\v x_{i,j}$ is a vector of observed characteristics and contextual variables 
specific to the pair, and 
$\bbeta$ is an unknown regression parameter to be estimated. 
In particular, data analysis based on  both the ERGM models and the 
 the variety of latent variable models mentioned above 
allow for 
estimation of such regression terms from complete, 
fully observed network data. 

In this article, we develop a type of likelihood that 
accommodates the ranked and censored nature of data from 
fixed rank nomination surveys, and allows for 
estimation of the type of regression effects described above. 
Additionally, we show that the failure to account for the 
censoring in such data can lead to biased inferences for 
certain types of regression effects, in particular, the 
effects of any characteristics specific to the nominators of the 
relations. 
In the next section, we introduce the fixed rank nomination (FRN) likelihood,
which accommodates  both the
ranked nature of FRN data and   the constraint 
on the number of nominations.
We relate this likelihood to two other likelihood functions that are in
use or may be appropriate for related types of network data collection
schemes: a likelihood based on ranks, and 
%the commonly used
a likelihood based on a 
probit  model, 
appropriate for unranked, uncensored binary network data. 
In Section 3
we provide both an  
analytical argument  and a simulation study
 that suggests that this latter ``binary'' likelihood
may  provide reasonable inference 
for some types of model  parameters, 
but misleading inference for others, in particular, those that estimate 
the effects of the nominators characteristics on network relations. 
This is further illustrated in an analysis of several 
adolescent social networks from the AddHealth study, in which we 
model the friendship preferences of students in the study as a function of 
individual and pair-specific explanatory variables
based on characteristics such as grade, grade point average, 
ethnicity and smoking and drinking behavior. 
%These data exhibit additional forms of incompleteness that 
%are addressed in the data analysis. 
A discussion follows in Section 5.

%% other way - understanding relat between X and Y 
%% Ad health 
%% FRNom
%% gap in existing methd

\section{Likelihoods based on fixed rank nomination data}
In this section we develop a 
type of 
likelihood function that is appropriate 
for modeling data that come from 
fixed rank nomination surveys.  
The likelihood is derived 
by positing a relationship
between the observed relational data $\m S$  and some underlying relational 
data $\m Y$ that the ranks are representing. 
In some situations,
% there is a well-defined  but unknown sociomatrix
%$\m Y$ that the matrix of ranks is representing. 
such a $\m Y$ is reasonably well defined. 
For example, 
the $i$th row of $\m S$ may record the 
top email recipients for individual $i$. 
In this case, the observed data $\m S$ is a coarsened version 
of the  sociomatrix $\m Y$ of email counts. 
In other situations a
definition of $\m Y$ is less precise, as with surveys that ask
people to nominate their ``top five friends.''
% but do not 
%define these 
%being more specific.
For either case, the likelihood developed below provides a statistical 
model for the ranked nomination data that  makes full use of the 
rank information and  accounts for the constraint on the 
number of nominations that may be made.
We contrast this likelihood to other likelihood  functions that 
do not make use of the rank data and/or do not account for the 
nomination constraint.

\subsection{Set-based likelihoods for ranked nomination data}
Let $\m Y=\{ y_{i,j}: i\neq j \}$ 
denote a sociomatrix of ordinal relationships 
among a population of 
$n$ individuals, so that $y_{i,j}>y_{i,k}$ means that 
person $i$'s relationship to person $j$ 
is in some sense % stronger or has more value 
stronger, of more value, or 
larger in magnitude
than their 
relationship to person $k$. 
Observation of $\m Y$ would allow for an analysis of the relationship 
patterns in the population, perhaps via a statistical model 
$\{ p(\m Y|\theta) : \theta\in \Theta\}$, where $\theta$ is an 
unknown parameter to be estimated.  
%For example, in the following sections 
%we consider a random-effects regression model for $\m Y$ of the form  
%\[ y_{i,j} = \bbeta^T \v x_{i,j} + a_i + b_j +\epsilon_{i,j}, \]
%$\v x_{i,j}$ is a vector 
%where $a_i$ and $b_j$ are 
 
As discussed in the Introduction, 
many surveys of social relations record only  incomplete 
representations of such a sociomatrix $\m Y$. 
%As discussed above, 
In positive  fixed rank nomination schemes, 
each individual provides 
an ordered ranking of 
people with whom they have a ``positive'' relationship, up to some limited 
number, say $m$.
One representation of such data is as a sociomatrix of 
scores $\m S = \{ s_{i,j} : i\neq j\}$, coded so that 
$s_{i,j}=0$ if $j$ is not nominated by $i$, 
$s_{i,j}=1$ if $j$ is $i$'s least favored nomination, and 
so on. Under this coding, 
$s_{i,j}> s_{i,k}$ if 
$i$ scores $j$ ``more highly'' than $k$, or if 
 $i$ nominates $j$ but not $k$.
Letting $a_i = \{1,\ldots, n\}\setminus \{i\}$ be
the set of individuals whom 
 person $i$ may potentially nominate,  
% (i.e.\ person $i$'s potential ``alters''), 
%and $d_i = \sum_{j\in \{ 1,\ldots, n\} \smallsetminus\{i\}   } 1(r_{i,j}>0 )$ be the number of 
%people ranked by $i$ (the outdegree of $i$), 
each observed outdegree $d_i \equiv\sum_{j\in a_i } 1(s_{i,j}>0 ) $ 
%the sociomatrix $\m S$ satisfies  
satisfies $d_i\leq m$. 
% for each row $i$ of the sociomatrix. 
%$ d_i \equiv\sum_{j\in a_i } 1(s_{i,j}>0 )\leq m$ for each row $i$. 

In order to make inference about $\theta$ from the observed 
scores $\m S$, 
the relationship between  $\m S$ and the unobserved 
relations $\m Y$ must be specified. 
The scores defined as above can be viewed as a coarsened and censored function
of the ordinal relations $\m Y$, or in other words, the sociomatrix $\m S$ 
is a many-to-one function of $\m Y$. 
The entries of $\m S$ can be written as an explicit function of 
$\m Y$ as follows:
\begin{equation}
 s_{i,j} = [ ( m -  {\rm rank}_i(y_{i,j} ) +1 ) \wedge 0 ] 
 \times 1( y_{i,j}>0 ) ,  
\label{eq:frnt}
\end{equation}
% rank here defined the opposite of ranks R
where ${\rm rank}_i(y_{i,j})$  is the rank of $y_{i,j}$ among 
the values in the $i$th row of $\m Y$, from high to low. 
Alternatively, some intuition can be gained by 
describing this 
function in terms of its inverse, defined by the following three 
associations:
\begin{eqnarray}
s_{i,j}> 0 &\Rightarrow & y_{i,j}>0  \label{eq:porc}\\
s_{i,j}  > s_{i,k} & \Rightarrow &  y_{i,j} > y_{i,k} 
  \label{eq:rnkc}\\ 
s_{i,j} = 0  \ \mbox{and} \  
 d_i < m &\Rightarrow & 
    y_{i,j}\leq 0  \label{eq:degc}. 
\end{eqnarray}
The first association follows from the definition of ranked individuals
as those with whom there is a positive relationship.
The second association follows from $\{ s_{i,j}:  j\in a_i \}$, the 
elements in the 
$i$th row of $\m S$, having the same order as 
$\{ y_{i,j} : j\in a_i\}$, the elements in the $i$th row of $\m Y$. 
The third association is a result of the censoring 
of the ranks:
If person $i$ did not nominate person $j$ ($s_{i,j}=0$) but could have ($d_i<m$), 
then their relationship to $j$ is not positive ($y_{i,j}<0$).
%, and so  
%$s_{i,j}=0$ and $m_i<m$ together imply $y_{i,j}\leq 0$. 
On the other hand, if 
$d_i = m$ then person $i$'s unranked relationships 
are censored, and so $y_{i,j}$ could be positive 
even though 
$s_{i,j}=0$.  In this case, all that is known about 
$y_{i,j}$ is that it is less than $y_{i,k}$ for any person 
$k$ that is ranked by $i$. 

Given a statistical model 
$\{ p(\m Y| \theta) : \theta\in \Theta \}$ 
for the underlying social relations $\m Y$, inference 
for the parameter $\theta$ can be based on a likelihood 
derived from the observed scores $\m S$. 
The likelihood is, as usual, the probability of the 
observed data $\m S$  as a function of the parameter $\theta$. 
To obtain this probability, 
let  $F(\m S)$ denote the set of $\m Y$-values that are
consistent with $\m S$ in terms of associations
(\ref{eq:porc}) - (\ref{eq:degc}) above. 
Since  the entries of $\m S$ are the observed scores if and only if 
$\m Y \in F(\m S)$, 
the likelihood is given by 
\[   
  L_F(\theta:\m S) = 
   \Pr(\m Y \in F(\m S)  | \theta)  =
 \int_{F(\m S)} p( \m Y | \theta) \ d\mu(\m Y), \]
where $\mu$ is a measure that dominates the probability densities
$\{ p(\m Y|\theta) : \theta\in \Theta \}$. 
We refer to a likelihood of this form, based on a set $F(\m S)$ 
defined by (\ref{eq:porc}) - (\ref{eq:degc}), 
   as a  fixed ranked nomination  (FRN)
likelihood, as it is
%appropriate 
%for analysis of data from positive fixed rank nomination surveys. 
derived from the probability distribution of the data 
obtained from a fixed rank nomination survey design. 

The FRN likelihood 
can be related to other likelihood functions that are used 
for ordinal  or binary  data. For example, 
consider the set  
$R(\m S) = \{ \m Y :  s_{i,j} > s_{i,k} \Rightarrow
  y_{i,j} > y_{i,k} \}$, defined by association 
 (\ref{eq:rnkc}) alone. The likelihood given by 
   $L_R(\theta:\m S) = \Pr(\m Y \in R(\m S)  | \theta)$ is known as a
rank likelihood for ordinal data, 
variants of which
have been
used for semiparametric regression modeling \citep{pettitt_1982} and 
 copula estimation   \citep{hoff_2007a}. 
Use of a
rank  likelihood for fixed rank nomination data is valid in some  sense, 
but not fully informative: In general we will have 
$F(\m S)  \subsetneq R(\m S)$, 
as the  information about $\m Y$ that $R(\m S)$ provides 
incorporates only one of the 
three conditions that defines $F(\m S)$.
This rank likelihood thus uses accurate but incomplete 
information about the value of $\m Y$, as compared to the information 
used by the FRN likelihood.

% in the sense that it makes use of some but not all 
%of the information about $\m Y$ provided by the ranks $\m R$. 
Another type of likelihood  often used to analyze relational data is 
obtained by relating $\m S$ and $\m Y$ as follows:
\begin{eqnarray}
s_{i,j} > 0  &\Rightarrow & y_{i,j}>0  \setcounter{equation}{2}  \\
s_{i,j} = 0  &\Rightarrow & y_{i,j}\leq 0 \setcounter{equation}{5}  . 
\end{eqnarray}
Letting $B(\m S) = \{ \m Y : s_{i,j} > 0  \Rightarrow  y_{i,j}>0, 
    s_{i,j} = 0  \Rightarrow  y_{i,j}\leq 0 \}$, 
the corresponding likelihood is given by 
$L_B(\theta:\m S) = \Pr(\m Y \in B(\m S) | \theta)$. 
We refer to this  as a binary likelihood, as 
probit and logistic models of binary relational data use this 
type of likelihood. 
To see this, note that the set $B(\m S)$ contains information 
only on the presence ($s_{i,j}>0$) or absence ($s_{i,j}=0$) of a
ranked relationship. As with probit or logit models, the presence or 
absence of a relationship corresponds to a latent variable (here $y_{i,j}$) 
being above or below some threshold (zero).
%, i.e.\  $y_{i,j}$  being 
%greater than or less than zero. 
Such a likelihood for fixed rank nomination data 
is neither fully informative nor valid: 
Not only  does it discard the information 
that differentiates among the ranked individuals, it also 
ignores the censoring on the outdegrees that results 
from the restriction on the number on individuals any one person 
may nominate.  In particular, 
$F(\m S)  \not\subset B(\m S)$ generally, and so the binary 
likelihood %given by $\Pr( \m Y \in B(\m R) | \theta)$ 
%is calculated from an integral partly over 
%$\m Y$-values that 
%could not have occurred, given the observed scores $\m S$. 
%In other words, the binary likelihood 
is based on the 
probability of the event $\{ \m Y \in  B(\m S)\}$,  subsets of  
which we know could not have occurred.

We note that the information about $\m Y$ provided by $\m S$ 
via equations (2) and (5) corresponds to the commonly used
representation of a relational dataset as an  edge list, i.e.\ 
a list of pairs of individuals between which there is an observed 
relationship. Such a representation ignores any information in the ranks, 
ignores the possibility of missing data and 
does not by itself convey any information about censored 
relationships.

\subsection{Bayesian estimation with set-based likelihoods}
The FRN, rank and binary likelihoods  can each be expressed as
the integral  of $p(\m Y|\theta)$ over a high-dimensional and
somewhat complicated set of $\m Y$-values, given by
$F(\m S)$, $R(\m S)$ and $B(\m S)$ for the three likelihoods respectively. Although such an integral will generally be intractable,
inference for $\theta$ can proceed using a Markov chain Monte
Carlo (MCMC) approximation to a Bayesian posterior distribution.
Given the observed ranks $\m S$ and
a prior distribution $p(\theta)$ over  the parameter space $\Theta$,
the joint posterior distribution  with
density $p(\theta,\m Y |\m S)$ can be approximated by
generating a Markov chain whose stationary distribution is
that of $(\theta ,\m Y)$ given $\m Y \in F(\m S)$,  $R(\m S)$ or  $B(\m S)$, 
depending on the likelihood being used. The values of
$\theta$ simulated from this chain provide an  approximation
to the (marginal) posterior  distribution of $\theta$ given 
the information from $\m S$.
One such MCMC algorithm
is the Gibbs sampler, which
iteratively simulates values of $\theta$ and $\m Y$
from their full conditional distributions.
Below we provide
Gibbs samplers for the FRN, binary and rank likelihoods
that can be used with any model 
for $\m Y$
that allows for simulation of each $y_{i,j}$ from 
$p(y_{i,j}|\theta,\m Y_{-(i,j)})$ constrained to an interval , 
where $\m Y_{-(i,j)}$ denotes the entries of $\m Y$ other than 
$y_{i,j}$. 
If simulation from this distribution is not available, 
the algorithms below can be modified by replacing such simulations 
with Metropolis-Hastings sampling schemes. 

%for $\m Y$ in which dyads are conditionally independent
%given a parameter $\theta$, i.e.\
%\[ p(\m Y | \theta ) = \prod_{i<j} p(y_{i,j},y_{j,i} | \theta). \]
%This includes a wide variety of relational models
%used in practice,
%as $\theta$ can be high-dimensional and include node-specific
%random effects and latent variables.

Given current values of $(\theta, \m Y)$, one step of a Gibbs sampler
for the FRN likelihood proceeds by updating the values as follows:
\begin{enumerate}
\item Simulate $\theta \sim p(\theta | \m Y)$. 
\item For each $i\neq j$, simulate
      $y_{i,j} \sim p( y_{i,j} |\theta, \m Y_{-(i,j)}, \m Y\in F(\m S) )$ as follows:
\begin{enumerate}
\item if $s_{i,j} >0$
simulate 
%$y_{i,j} \sim  p(y_{i,j} | y_{j,i}, \theta)\times 
%1 (  \max \{ y_{i,k}: s_{i,k}<s_{i,j}  \} \leq  y_{i,j} \leq  \min \{ y_{i,k}: s_{i,k}>0 \}   );   $
\[ y_{i,j} \sim  p(y_{i,j} | \m Y_{-(i,j)}, \theta)\times 
1 (  \max \{ y_{i,k}: s_{i,k}<s_{i,j}  \} \leq  y_{i,j} \leq  \min \{ y_{i,k}: s_{i,k}>s_{i,j} \}   ); \]
\item if $s_{i,j} =0$ and $d_i<m$ ,
   simulate $y_{i,j} \sim  p(y_{i,j} | \m Y_{-(i,j)}, \theta )\times 
1 ( y_{i,j} \leq 0);$
\item
 if $s_{i,j} =0$ and $d_i=m$ ,
   simulate $y_{i,j} \sim  p(y_{i,j} | \m Y_{-(i,j)}, \theta )\times 
1 ( y_{i,j} \leq  \min \{ y_{i,k}: s_{i,k}>0 \}   ).$
\end{enumerate}
\end{enumerate}
In the above steps, ``$y\sim f(y)$'' means ``simulate $y$ from a distribution 
with density proportional to $f(y)$''. 
For each ordered pair $(i,j)$, step 2 of this algorithm will
generate a value of $y_{i,j}$ from its full conditional distribution,
constrained so that conditions (2)-(4) that define the FRN likelihood are met.
Gibbs samplers for the binary and rank likelihoods are obtained
by replacing step 2 of the above algorithm with
different
constrained simulation schemes. For the binary likelihood, step 2 becomes
\begin{enumerate}
\item[2.] For each $i\neq j$, simulate
      $y_{i,j} \sim p( y_{i,j} |\theta, \m Y_{-(i,j)}, \m Y\in B(\m S) )$ as follows:
\begin{enumerate}
\item if $s_{i,j} >0$
simulate $y_{i,j} \sim  p(y_{i,j} | \m Y_{-(i,j)}, \theta )\times 
1 (  y_{i,j}>0   );   $
\item if  $s_{i,j} =0$
simulate $y_{i,j} \sim  p(y_{i,j} | \m Y_{-(i,j)}, \theta )\times 1 (  y_{i,j}\leq0   ).$
\end{enumerate}
\end{enumerate}
For the rank likelihood, the corresponding step is
\begin{enumerate}
\item[2.] For each $i\neq j$, simulate
      $y_{i,j} \sim p( y_{i,j} |\theta, \m Y_{-(i,j)}, \m Y \in R(\m S) )$ as
%\begin{itemize}
%\item 
\[ y_{i,j} \sim  p(y_{i,j} | \m Y_{-(i,j)}, \theta )\times 
1 (  \max \{ y_{i,k}: s_{i,k}<s_{i,j}  \} \leq  y_{i,j} \leq  \min \{ y_{i,k}: s_{i,k}> s_{i,j } \}   ) . \]
%\end{itemize}
\end{enumerate}

It is also straightforward
to extend this Gibbs sampler
to accommodate certain types of missing data.
For example, some students participating in
the AddHealth study were not included on their school's roster of possible
nominations. In this case, $s_{i,j}$ is missing for each unlisted
student $j$ and every other student $i$.
If the relations $\{y_{i,j}: i\in \{1,\ldots, n\}\setminus{j}\}$
of the students to a particular student $j$ are independent
of whether or not student $j$ is on the roster, the
observed rankings $\m S$ provide no information about
 $\{y_{i,j}: i\in \{1,\ldots, n\}\setminus{j}\}$
and thus the full conditional distribution of $y_{i,j}$
is $p(y_{i,j} | y_{j,i} ,\theta)$, unconstrained.
Simulating $y_{i,j}$ from this distribution for each missing
$s_{i,j}$ allows for imputation of friendship  nominations to
unlisted students, and generally facilitates simulation of the parameter
$\theta$  in the MCMC algorithm.

The above algorithms, or simple variants of them, are straightforward to 
implement for many statistical models of social networks and 
relational data.
For example, latent variable models based on 
conditional dyadic independence 
(such as those used in
\citet{nowicki_snijders_2001} and \citet{hoff_2005a})
will satisfy $p(y_{i,j} | \m Y_{-(i,j)} ,\theta) = 
  p(y_{i,j} | y_{j,i} ,\theta)$, which makes step 2 of the above algorithms
much easier. Additionally, in these models the unobserved relations
$\m Y$ can be taken to be normally distributed, and so 
step 2 involves simulations from constrained normal distributions, 
which are fairly easy to implement. 
Furthermore, it is not necessary in step 1 that 
$\theta$ is simulated from its full conditional distribution. 
Instead, all that is necessary is that it can be simulated in a way 
that makes the stationary distribution of the Markov chain 
equal to the posterior distribution $p(\theta, \m Y|\m S)$. 
This can be achieved with a block Gibbs sampler for different 
components of $\theta$, or with some other Metropolis-Hastings update.

\section{Comparing likelihoods in social relations regression models} 
While we have argued that $L_R(\theta:\m S)$ and
$L_B(\theta:\m S)$ may be inappropriate likelihoods
for estimating $\theta$ from fixed rank nomination data, in practice they may
provide inference that  approximates that obtained
from $L_F(\theta:\m S)$, at least for some aspects of $\theta$ or 
under certain conditions. 
To explore this possibility, we consider inference under the different 
likelihoods in the 
case where $\theta$ represents the parameters in the following standard 
regression  model for relational data:
\begin{align} 
y_{i,j}  & = \bbeta^T \v x_{i,j}  + a_i + b_j + \epsilon_{i,j} 
\label{eq:srm} \\
%\left ( (\begin{smallmatrix} a_i \\ b_i \end{smallmatrix}), i=1,\ldots, n \right) & \sim 
% \mbox{ i.i.d.\ normal}(\v 0, \Sigma_{ab} ) \nonumber  \\ 
\left( ( \begin{smallmatrix} \epsilon_{i,j}  \\ \epsilon_{j,i} \end{smallmatrix} ) , i\neq j \right) & \sim 
 \mbox{ i.i.d.\ normal}(\v 0,\sigma^2 (\begin{smallmatrix} 1 & \rho \\ \rho & 1 \end{smallmatrix} )  )  \nonumber
\end{align}
The additive row effect $a_i$ is  often interpreted
as a measure of
person $i$'s ``sociability,'' whereas the additive column
effect $b_i$ is taken as a  measure of $i$'s  ``popularity.'' 
The parameter $\rho$  represents 
potential correlation between
$y_{i,j}$ and $y_{j,i}$. 
In a mixed-effects version of this model, 
the possibility that 
a person's sociability $a_i$ is correlated with their popularity $b_i$ 
can be represented with a 
 covariance matrix $\Sigma_{ab}$.  
The covariance among the elements of $\m Y =\{ y_{i,j} : i\neq j\}$ induced 
by $\Sigma_{ab}$ and $\Sigma_\epsilon = \sigma^2 (\begin{smallmatrix} 
   \rho & 1 \\ 1 & \rho \end{smallmatrix} ) $ 
       is called the social relations 
model \citep{warner_kenny_stoto_1979}, and has been frequently used as a model 
for 
continuous relational data \citep{wong_1982, gill_swartz_2001, li_loken_2002} 
as well as a component of a generalized linear model
for binary or discrete network data \citep{hoff_2005a}. 
This model is very similar to the ``$p2$'' model  
of \citet{vanduijn_snijders_zijlstra_2004}, which is an  extension of the  
well-known log-linear $p1$ model of  \citet{holland_leinhardt_1981}. 
Like the  social relations model, the $p2$ model 
has row- and column-specific random effects and allows the 
network relationships to depend on regressors. 
We note that these models cannot represent commonly observed network 
patterns such as transitivity, clustering or stochastic equivalence, 
unless these patterns can be captured by covariate effects. 
However, the set-based likelihoods  presented above can be applied 
to network models that do account for such patterns, as will be discussed in 
Section 5.

In what follows, we consider estimation of the parameters 
in model \eqref{eq:srm}
for the underlying relations $\m Y$, when the observed data include
only the censored nomination scores $\m S$, given by \eqref{eq:frnt}.
As we will show,
the likelihoods  $L_R(\theta:\m S)$ and $L_B(\theta:\m S)$
are  inappropriate for estimation of any row-specific effects, 
i.e.\ terms in the
regression model \eqref{eq:srm} that are constant across the row
index $i$, the index of the nominators of the relations. 
%This includes any regressors that are row-specific, as well 
%as row-specific random effects.  
This limitation includes any nominator-specific regressors, as well as 
nominator-specific random effects. 
We first show this analytically, and then confirm the results with
a small simulation study.
In contrast to the case for row effects, 
the simulation study suggests that the binary and rank 
likelihoods may provide reasonable inference for column-specific 
effects and certain types of 
dyad-specific effects.

\subsection{Estimation of additive row effects}
In assessing the ability of $L_R(\theta:\m S)$ and $L_B(\theta:\m S)$ to estimate row-specific 
effects, it 
will be convenient to reparameterize  
the model to separate these
 terms out from the rest.  
We rewrite \eqref{eq:srm} as
\begin{align*} y_{i,j} & = \alpha_i + \v \beta_{cd}^T \v x_{i,j}+\epsilon_{i,j} \\
  \alpha_i & = \bbeta_r^T \v x_i + a_i 
\end{align*}
so that 
$\alpha_i$ is equal to $a_i$ from \eqref{eq:srm} plus any 
regression effects $\v\beta_r^T\v x_i$ that are constant across rows (i.e., are based on any characteristics of the nominator of the tie), and $ \bbeta_{cd}^T \v x_{i,j}$ now represents 
any other column-specific or dyad-specific regression terms, including the additive column effect $b_j$. 

We first show that the row-specific effects $\v \alpha= (\alpha_1,\ldots, \alpha_n)$
are not estimable using the rank likelihood $L_R(\theta:\m S)$.
Recall that the rank likelihood is given by 
\begin{align*} 
L_R(\theta : \m S)& = \Pr( \m Y \in R(\m S) | \theta ) \\ 
%R(\m S) & =\{ \m Y  :  y_{i,j}> y_{i,k} \ \forall \{i,j,k\} : s_{i,j} > s_{i,k} \} 
(\m S) & =\{ \m Y  :  y_{i,j}> y_{i,k}  \mbox{ for all }  \{i,j,k\} \mbox{ such that }  s_{i,j} > s_{i,k} \} 
\end{align*}
where here we take $\theta=\{ \v \alpha, \v \beta , \sigma^2,\rho \}$. 
For a given row $i$, 
the likelihood only provides information on the relative ordering 
of the $y_{i,j}$'s, and not their overall magnitude. 
To see why this precludes estimating row effects, 
note that the 
ordering of the $y_{i,j}$'s within row $i$ is unchanged 
by the addition of a constant, and so if
$\m Y \in R(\m S)$, so is $\m Y + \m c \m 1^T $ for 
any vector $\m c$ in $\mathbb R^n$.  
Therefore, 
\begin{align*}
\Pr( \m Y \in R(\m S) | \v \alpha ,\bbeta_{cd},\Sigma_{\epsilon} ) &=
\Pr( \m Y + \m c\m 1^T \in R(\m S) | \v \alpha ,\bbeta_{cd},\Sigma_{\epsilon} ) \\
&= \Pr( \m Y  \in R(\m S) | \v \alpha + \m c ,\bbeta_{cd},\Sigma_{\epsilon} )
\end{align*}
for all $\m c\in \mathbb R^n$, and so 
$\Pr( \m Y \in R(\m S) | \v \alpha ,\bbeta_{cd},\Sigma_{\epsilon} )$ 
cannot be a function of $\v \alpha$, and therefore cannot be used to 
estimate $\v \alpha$.

%For example, 
%letting $l_F(\theta) = \log L_F(\theta:\m S)$
%and $l_R(\theta) = \log L_R(\theta:\m S)$ be the log-likelihoods,
%we have
%\begin{eqnarray*}
%l_F(\theta:\m S) &= &\log \Pr(\m Y \in F(\m S) | \theta)  \\
% &=& \log \left \{   \Pr( \m Y\in R(\m S) | \theta ) \times 
%    \Pr( \m Y\in F(\m S) |\m Y\in R(\m S) , \theta )  \right \} \\
% &=& l_R(\theta:\m S) + \log  \Pr( \m Y\in F(\m S) |\m Y\in R(\m S),\theta ). 
%\end{eqnarray*}
%The difference between the two log-likelihoods is given by the
%log probability of $F(\m S)$ given $R(\m S)$.  If this probability
%is small relative to  $\Pr(\m Y\in  R(\m S) | \theta)$, or
%only depends weakly on $\theta$, then the two likelihoods will
%be approximately the same.

Estimation of row effects is also problematic for the 
the binomial likelihood $L_B(\theta:\m S)$. 
Recall that the binomial likelihood is given by 
\begin{align*} 
L_B(\theta : \m S)& = \Pr( \m Y \in B(\m S) | \theta ) \\ 
%B(\m S) & =\{ \m Y  :  y_{i,j}> 0 \ \forall \{i,j\} : s_{i,j} > 0 ,  \ \mbox{and} \  y_{i,j}< 0 \ \forall \{i,j\} : s_{i,j} = 0 \}.
B(\m S) & =\{ \m Y  :  y_{i,j}> 0 \ \mbox{ for all }  \{i,j\} \mbox{ such that  }  s_{i,j} > 0 ,  \  y_{i,j}< 0 \ \mbox{ for all } \{i,j\} \mbox{ such that } s_{i,j} = 0 \}.
\end{align*}
Under the binomial likelihood, 
the data are essentially assumed to be coming from a 
probit regression model and 
the estimate of the row effect $\alpha_i$ is largely 
determined by the 
number of nominations that person $i$ will make, i.e.\ their 
observed outdegree $d_i$. 
This is appropriate in the absence of censoring, where $d_i$
reflects the number of positive relations that person $i$ has. 
However, in the presence of censoring a person's outdegree 
(and therefore their estimated row effect) 
may be controlled by the 
the maximum number of nominations $m$ they are allowed to make. 
For example, consider a person $i$ 
having many more positive relations, say $\tilde d_i$, than the 
number of allowed nominations $m$.  
%If the number of individuals in the network $n$ is much larger than 
%$m$, then 
In this case, $d_i$ will equal $m$ and  the binomial likelihood will underestimate $\alpha_i$, 
reflecting the observed outdegree of $m$ 
rather than 
person $i$'s actual outdegree $\tilde d_i$. 
Additionally, if $\tilde d_i$ is much higher than $m$ for many individuals, 
then many individuals will make the maximum number of nominations and 
the variability in the observed outdegree will be low. 
Inference under the binomial likelihood will incorrectly attribute
this to low 
variability among the $\alpha_i$'s. 
As one component of the variability in the $\alpha_i$'s is 
the variability in the row-specific regression effects $\bbeta_r^T \v x_i$,
underestimated variability among the $\alpha_i$'s will translates into 
underestimates of the magnitude of $\bbeta_r$.

%This is not to say that we should expect the FRN likelihood to infer the
%true activity level $\tilde m_i/n$. Rather, the FRN likelihood
%should recognize the censoring, and thus reflect this
%additional uncertainty in $\alpha_i$.

We make this argument more concrete via an analytic comparison 
between the binomial  and FRN likelihoods, 
showing that the 
binomial likelihood 
for the social relations model 
is approximately equal to the 
FRN likelihood for a model with no row-specific variability. 
%This result implies that, 
%since the 
%the latter likelihood essentially estimates row-specific 
%variability as being zero, the former likelihood will underestimate this 
%variability. 
For simplicity, we compare likelihoods based on the ranked nomination 
data 
from a single nominator who makes $m$ nominations. 
Denote this individual's unobserved relations
to the other $n-1$ individuals 
as $\v y  =\{y_{j}: j=1,\ldots, n-1\}$, and the observed nomination scores 
as  $\v s =\{s_{j}: j=1,\ldots, n-1\}$. 
From equations (2)--(4), 
the FRN likelihood is 
%$L(\theta:\v s) = \Pr( A(\v s) \cap B(\v s) | \theta)$, 
the joint probability of the events
$A(\v s)  = \{ \v y\in \mathbb R^{m-1} :  y_{(m)} >  \cdots > y_{(1)} >0  \} $ 
and 
$B(\v s) = \{ \v y\in \mathbb R^{m-1}: y_{(1)} > \max\{ y_{j} : s_j =0  \}\}$,  
where $y_{(k)} $ denotes the nominator's relationship to 
the person with the $k$th lowest non-zero score.
%Specifically, $L(\theta:\v s) = \Pr( A(\v s) \cap B(\v s) | \theta)$. 
%Let $A\subset \mathbb R^{m-1}$ and $B\subset \mathbb R^{m-1}$ denote
%the sets of $(y_1,\ldots, y_{m-1})$-values that satisfy these two inequalities,
%respectively.

Suppose we use this FRN likelihood with a model for $\v y$ where 
the $y_j$'s are independent with 
$y_{j}\sim N(\v\bbeta_{cd}^T\v x_j,1)$, and $\bbeta_{cd}^T\v x_j$ contains no 
intercept (this would correspond to a model with no row-specific 
effects, when extended to a likelihood based on 
data from multiple nominators). 
Letting $\phi$ and $\Phi$ be the standard normal density  and CDF 
respectively, 
the no-intercept FRN likelihood can be expressed 
as 
\begin{align}
L_{F}(\bbeta_{cd}: \v s )  &= \Pr(  A(\v s) \cap B(\v s) |\bbeta_{cd}) \nonumber \\
&= \int_0^\infty \Pr(A(\v s) \cap B(\v s)  | \bbeta_{cd}, y_{(1)} ) \times 
    \phi(y_{(1)} - \bbeta_{cd}^T \v x_{(1)} ) \ dy_{(1)} \nonumber  \\ 
&= \int_0^\infty \Pr(A(\v s) | \bbeta_{cd}, y_{(1)} ) \Pr(B(\v s) | \bbeta_{cd}, y_{(1)} ) \times 
    \phi(y_{(1)} - \bbeta_{cd}^T \v x_{(1)} ) \ dy_{(1)},
\label{eq:lf}
\end{align}
since $A(\v s)$ and $B(\v s)$ are conditionally independent given $y_{(1)}$. 
Now $\Pr(B(\v s)  | \bbeta_{cd}, y_{(1)} )$ is given by
\begin{equation}
 \Pr(B(\v s) | \bbeta_{cd}, y_{(1)} ) = 
\prod_{j:s_j=0} \Pr(y_j<y_{(1)} | \bbeta_{cd} , y_{(1)} ) 
%\prod_{j:s_j=0} \Phi ( y_{(1)} - \bbeta_{cd}^T\v x_j) \\ 
 = \prod_{j:s_j=0} [1-\Phi (  \bbeta_{cd}^T\v x_j - y_{(1)}) ], 
\label{eq:pb}
\end{equation}
which is the same as the contribution of the ``zeros'' to a 
probit likelihood for binary data with linear predictor 
$\alpha  +\bbeta_{cd}^T\v x_j$, where $\alpha= -y_{(1)}$. 

Using Bayes' rule, we can write  $\Pr(A(\v s) | \bbeta_{cd},y_{(1)})$ as
\begin{align}
 \Pr(  A(\v s) | \bbeta_{cd}, y_{(1)})   &= 
 \left ( \prod_{j=2}^m \Pr(  y_{(j)}>y_{(1)}  |\bbeta_{cd},y_{(1)})  \right) \times  \label{eq:pa} \\
 & \qquad \Pr(   y_{(2)}< \cdots   <  y_{(m)} |y_{(1)} , \bbeta_{cd}, \{ y_{(1)}<y_{(j)},j=2,\ldots m \}) \nonumber  \\
 &\equiv \left ( \prod_{j=2}^m   \Phi( \bbeta_{cd}^T \v x_{(j)} - y_{(1)} )   \right)   
          \times h(y_{(1)},\bbeta_{cd}).  \nonumber
\end{align}
Note that the first term is equivalent to the contribution of the ``ones'' to a
probit likelihood for binary data with linear predictor
$\alpha  +\bbeta_{cd}^T\v x_j$, where $\alpha= -y_{(1)}$. 
%Here, $h(y_{(1)},\bbeta_{cd}: y_{(2)},\ldots, y_{(n)} )$ is the conditional probability
%of a particular ordering among a set of $m-1$ latent affinities,
%given that they are all above the common value $y_{(1)}$.
Combining  \eqref{eq:pb} and \eqref{eq:pa}  gives
\begin{align*}
\Pr( A(\v s) \cap B(\v s) | \bbeta_{cd},y_{(1)}=-\alpha ) &=  
   \prod_{j=2}^m   \Phi( \alpha+\bbeta_{cd}^T \v x_{(j)}  )  \times 
 \prod_{j:s_j=0} [1-\Phi ( \alpha+  \bbeta_{cd}^T\v x_j ) ]  \times 
    h(-\alpha,\bbeta_{cd}) \\ 
 &= 
\left ( 
   \prod_{j:s_j\neq 1}
   [1- \Phi ( \alpha+ \bbeta_{cd}^T\v x_j )]^{(s_j=0)} 
   \Phi ( \alpha+ \bbeta_{cd}^T\v x_j ) ^{(s_j>0)} 
   \right  )
   h(-\alpha,\bbeta_{cd} )  \\ 
&=
 L_B(\alpha,\bbeta_{cd} : s_{-(1)}) \times h(-\alpha,\bbeta_{cd} ).
\end{align*}
where $ L_B(\alpha,\bbeta_{cd} : s_{-(1)}) $
is
exactly the binomial likelihood,
absent information from the lowest ranked nomination,
under the probit  model  with linear predictor 
$\alpha + \bbeta_{cd}^T \v x_{j}$. 
%  in which the $\epsilon_j$'s
%are i.i.d.\ standard normal errors.  
Incorporating this expression into equation \eqref{eq:lf}
shows that relationship between
the no-intercept FRN likelihood and this binomial probit likelihood is 
\begin{align*} 
L_F(\bbeta_{cd}: \v s ) = \int L_B(\alpha,\bbeta_{cd}: \v s_{-(1)}  )
 \times  g(\alpha,\bbeta_{cd})  \ d\alpha, 
%[  h(-\mu,\bbeta_{cd})  \phi( \mu+ \bbeta_{cd}^T x_{(1)}) ] \ d\mu.
\end{align*} 
where $g(\alpha,\bbeta_{cd}) =1_{(-\infty,0)}(\alpha) h(-\alpha,\bbeta_{cd} )  \phi( \alpha+ \bbeta_{cd}^T x_{(1)})$. A Laplace approximation to this integral gives 
\[ \log L_F(\bbeta_{cd}: \v s ) \approx \log L_B(\hat \alpha,\bbeta_{cd}:\v s_{-(1)}   ) +
    \log g(\hat \alpha,\bbeta_{cd}) + c, \]
where $\hat \alpha$ is the maximizer in $\alpha$ of the integrand and $c$ 
does not depend on $\bbeta_{cd}$. 

Proceeding heuristically, 
we generally expect
$L_B(\alpha,\bbeta_{cd}: \v s_{-(1)} )$ to be close to
$L_B(\alpha,\bbeta_{cd}: \v s )$, the binary likelihood based on
the nominator's full set of  scores,
as the former is lacking only the information on one ranked individual.
Furthermore, if 
$n$ is much larger than $m$, then we expect that $g$
will be relatively flat as a function of
$(\alpha,\bbeta_{cd})$ as  compared to $L_B$, as the latter is a
probit
likelihood
based on $n >> m$ observations, and the former involves
the conditional
probability of a particular relative ordering among only  $m-1$
relations. As a result, the maximizer in $\alpha$ of 
$\log L_B(\alpha,\bbeta_{cd}:\v s ) + \log g(\alpha, \bbeta_{cd})$,
should be close 
to the maximizer  in $\alpha$ of $\log L_B(\alpha,\bbeta_{cd}:\v s )$, and 
  $\log g(\hat \alpha,\bbeta_{cd}) $ should be relatively flat as
 a function of $\bbeta_{cd}$ 
 compared to 
  $\log L_B(\hat \alpha,\bbeta_{cd})$. 
Combining these approximations suggests that 
\[ \log L_F(\bbeta_{cd}: \v s ) \approx \log L_B(\hat \alpha,\bbeta_{cd}:\v s) +d, \]
where  
$\hat \alpha$ is the maximizer in $\alpha$ of $ L_B(\hat \alpha,\bbeta_{cd}:\v s)$ and 
$d$ is (roughly) constant in $(\alpha,\bbeta_{cd})$. 

Extending this approximation to the case of ranked nominations from $n$ 
individuals, we have 
\begin{align*} 
\log L_F(\bbeta_{cd}: \m S )  & \approx \sum_{i=1}^n \log L_B(\hat \alpha_i,\bbeta_{cd}:\v s_i) +  k  \\
 &=\log  L_B(\hat \alpha_1,\ldots,\hat \alpha_n,\bbeta_{cd}:\m S) + k
\end{align*}
where for convenience we have ignored the possibility of dyadic correlation 
between $\epsilon_{i,j}$ and $\epsilon_{j,i}$. 
The result suggests that if most individuals make the maximum number of 
nominations, then the binomial likelihood with row-specific effects $\alpha_1,\ldots, \alpha_n$ should give roughly the same fit to the data 
as the FRN likelihood lacking any such effects. Estimation using the latter 
likelihood is equivalent to setting any row-specific regression 
coefficients 
$\v \beta_r$ to zero, and setting the across-row variance of any 
random effects $a_1,\ldots, a_n$ to zero as well. 
As the fit under the binomial likelihood will be similar, we expect it 
to provide underestimates of the magnitude of $\v\beta_r$ and the variance 
of the $a_i$'s.  However, these results do not preclude the possibility of 
approximately correct 
inference for 
column- and
dyad-specific effects, represented here by 
$\bbeta_{cd}$.

\subsection{Simulation study} 
We evaluated the above claims numerically with a small simulation study,
comparing parameter estimates for the social relations model \eqref{eq:srm}
obtained from the FRN, rank and binomial likelihoods. 
Specifically, we generated 
relational data $\m Y$ from an SRM  as in \eqref{eq:srm}
with random row and column effects: 
\begin{align} 
y_{i,j}  & = \bbeta^T \v x_{i,j}  + a_i + b_j + \epsilon_{i,j} 
\label{eq:srmab} \\
\left ( (\begin{smallmatrix} a_i \\ b_i \end{smallmatrix}), i=1,\ldots, n \right) & \sim 
 \mbox{ i.i.d.\ normal}(\v 0, \Sigma_{ab} ) \nonumber  \\ 
\left( ( \begin{smallmatrix} \epsilon_{i,j}  \\ \epsilon_{j,i} \end{smallmatrix} ) , i\neq j \right) & \sim 
 \mbox{ i.i.d.\ normal}(\v 0,\sigma^2 (\begin{smallmatrix} 1 & \rho \\ \rho & 1 \end{smallmatrix} )  ) ,  \nonumber
\end{align}
where our mean model had the following form:
\[ \v\beta^T \v x_{i,j}  =\beta_0 + \beta_r x_{i,r} + \beta_c x_{j,c} + \beta_{d_1} x_{i,j,1} + \beta_{d_2} x_{i,j,2}. \]
In this model, $x_{i,r}$ 
and $x_{j,c}$
are individual-level characteristics 
of person $i$ 
as a nominator and person $j$ as a nominee, respectively, 
and could quantify 
things such as smoking behavior  or grade point average of the individuals. 
The two dyad-specific characteristics 
$x_{i,j,1}$ and $x_{i,j,2}$ are specific to each pair of individuals, 
and could represent such things as 
the amount of time spent together, or an indicator of co-membership to a common group.  For each $\m Y$ generated from this model, we obtained 
the corresponding nomination scores under the FRN scheme using the 
relationship in Equation \ref{eq:frnt}.

We generated 16 networks of $n=100$ individuals each from this model, using the 
following parameter values:
\begin{itemize}
\item$\beta_0 = -3.26,\  \beta_r= \beta_c = \beta_{d_1} = \beta_{d_2} =1$;
\item$ \Sigma_{ab} =  (\begin{smallmatrix} 1 & .5 \\ .5 & 1 \end{smallmatrix} )   , \ \Sigma_{\epsilon} = (\begin{smallmatrix} 1 & .9 \\ .9& 1 \end{smallmatrix} )   $;
\item $\{ x_{1,r},\ldots, x_{n,r}\},
       \{ x_{1,c},\ldots, x_{n,c}\}, 
       \{ x_{i,j,1} : i\neq j\} \sim $ i.i.d.\ $N(0,1)$, 
\item  $x_{i,j,2} = z_{i}z_{j}/.42 $, where 
  $z_1,\ldots, z_n\sim$ i.i.d.\ binary($1/2$). 
\end{itemize}
The second dyadic characteristic  $x_{i,j,2}$ can be viewed as an 
indicator of co-membership to a common group,  of which each individual 
is a member with probability 1/2. The product $z_iz_j$ is divided 
by $0.42$ to give $x_{i,j,2}$ a standard deviation of 1, so that it is 
on the same scale as the other characteristics. 

The value of the intercept, $\beta_0= -3.26$, was chosen so that 
15\% of the $y_{i,j}$'s were greater 
than zero, making the average uncensored outdegree equal to 15. 
We generated 16 fixed rank nomination datasets under this scheme, 
8 for which the maximum number of nominations was $m=5$, and 8 for which 
this number was $m=15$. The former value resulted in a high degree 
of censoring, with 59\% of uncensored outdegrees $\tilde d_i$ being  greater 
than $m$. The censoring rate under $m=15$ was 38\%.

For each dataset, we obtained parameter estimates and confidence intervals 
under the FRN, binomial and rank likelihoods using an MCMC 
approach based on the procedure described in Section 2.2. 
%(more details are provided in the supplementary material). 
We ran the MCMC algorithms 
for 100,500 iterations, dropped the first 500 iterations for burn-in 
and saved parameter values every 25th iteration, resulting in 
4,000 simulated values of each parameter from which to make inference. 
Average effective sample sizes for the parameter estimates 
(an assessment of the MCMC approximation)
% presented below 
were 1080, 764 and 238  for the binomial, FRN  and rank likelihoods, 
respectively.

\begin{figure}
\centerline{\includegraphics[width=6.5in]{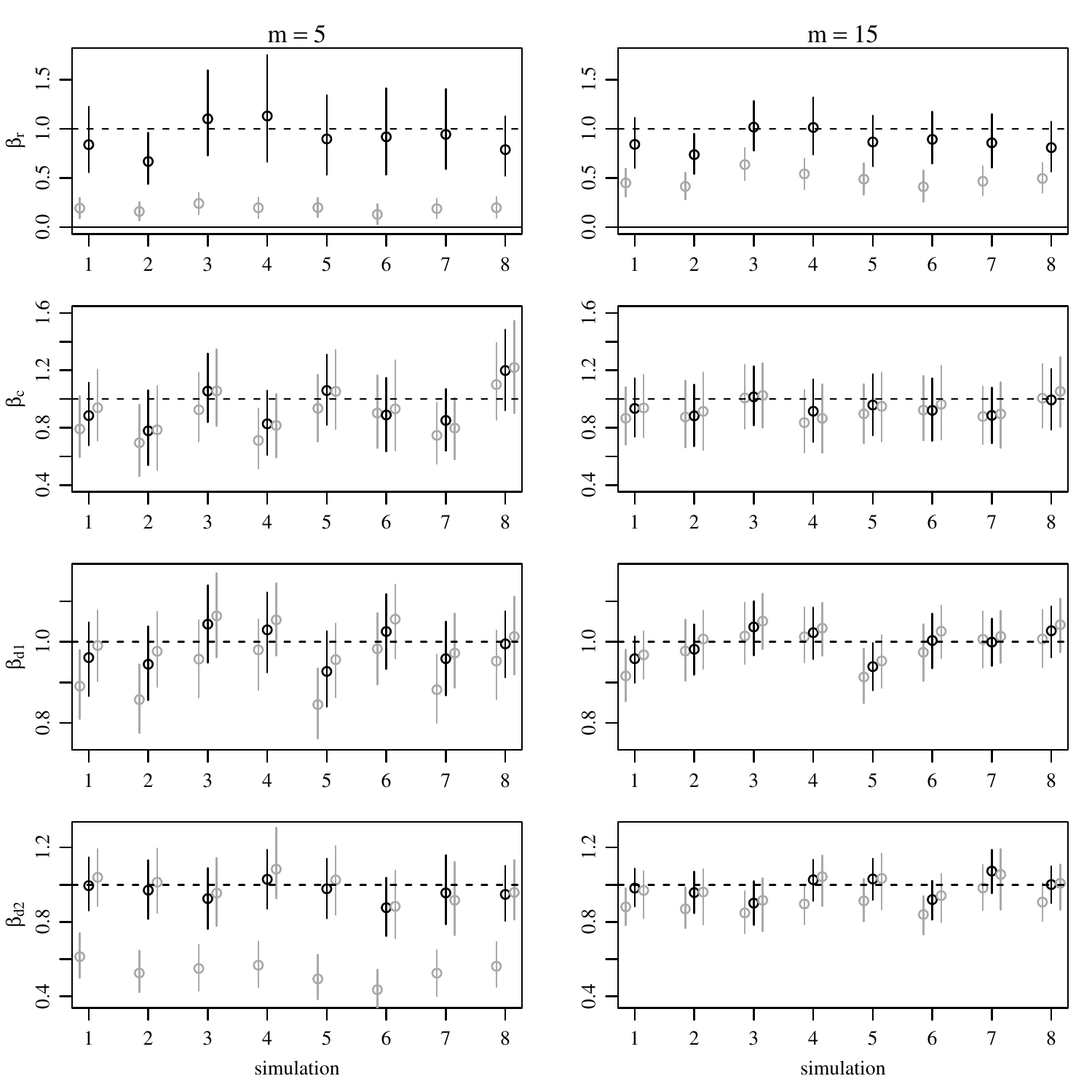}}
\caption{Confidence intervals under the three different likelihoods 
for the 16 simulated datasets. For each plot in the first  row, 
the confidence intervals are based on binomial and FRN likelihoods, 
from left to right. 
For each plot in the remaining rows, the groups of three confidence intervals are based on binomial, FRN and rank likelihoods, from left to right. }
\label{fig:simres}
\end{figure}

Figure \ref{fig:simres}  plots posterior 2.5\%, 50\% and 97.5\% 
quantiles for $\{\beta_r,\beta_c,\beta_{d_1},\beta_{d_2}\}$ 
across
all 16 datasets and for each of the likelihoods. 
These quantiles provide Bayesian point estimates
(the 50\% quantiles) and 95\% confidence intervals 
(the 2.5\% and 97.5\% quantiles). 
The  plot in the lower left-hand corner of the figure, for example, 
gives estimates and confidence intervals 
for $\beta_{d_2}$
when $m=5$, across the three likelihoods and eight 
simulated datasets. 
The intervals are plotted in groups of three, representing the 
binomial, FRN and rank likelihoods from left to right. 
The first row of the figure only gives intervals for 
the binomial and FRN likelihoods from left to right, as the rank likelihood 
cannot estimate any effects corresponding to regressors that 
are constant across  nominators. 

The results of these simulations are consistent with the preceding discussion 
of the inadequacies of the binomial likelihood.
For example, the first row of the figure highlights
potential problems with the binomial likelihood  in terms of 
estimating 
regression coefficients of nominator-specific regressors. 
Such  coefficients 
relate outdegree  heterogeneity 
to individual-specific effects. 
When the amount of censoring is large,  the heterogeneity of the 
censored outdegrees is low and so any nominator-specific regression 
coefficients will be erroneously estimated by the binomial likelihood 
as being low in magnitude as well. 
This degree of underestimation 
is reduced in the case $m=15$, but is still substantial. 
%For both values of $m$, the FRN likelihood appears to provide 
%reasonable parameter estimates and confidence intervals. 
Additionally, for both values of $m$, the confidence intervals for $\beta_r$ 
under the binomial likelihood are 
substantially narrower than those under the FRN likelihood. 
Taken together, these results indicate that inference under 
the binomial likelihood can lead not only to overconfident inference, 
but overconfidence in the wrong parameter values.

The second and third row of the plots in the figure suggest 
that the binomial likelihood estimates of column- and 
dyad-specific regression coefficients 
$\beta_c$ and $\beta_{d_1}$, while not as accurate as 
those from the FRN likelihood estimates, are not 
unreasonable. Similarly, the rank likelihood estimates
perform similarly to those obtained from the FRN likelihood. 
In contrast, the binomial likelihood estimates of $\beta_{d_2}$ 
perform quite 
poorly as compared to those from  the FRN or rank likelihood. 
The difference between estimation of $\beta_{d_2}$ and $\beta_{d_1}$
is that, unlike $\m X_1 = \{ x_{i,j,1} \} $, the matrix  $\m X_{2} = \{ x_{i,j,2}\}$ exhibits substantial 
row variability. Recall that $x_{i,j,2}=z_i z_j/.42$ 
is essentially the indicator of 
co-membership to a group. If individual $i$ is not in the group, then 
the $i$th row of $\m X_2$ is all zeros, whereas if they are in the group, 
then half the entries in the $i$th row are nonzero (as half of the 
population is in the group).  
By ignoring the censoring, the binomial likelihood underestimates 
the row variability in $\m Y$, and thus also the 
variability that can be attributed to the row variation in 
$\m X_2$. 

\subsection{Information in the ranks}
We have attributed the biases of the binomial likelihood estimators
to the fact that the binomial likelihood does not account for the 
censored nature of the data. However, it is fairly straightforward to 
modify the binomial likelihood to account for the censoring. 
Recall that the binomial likelihood was defined as 
$L_B(\theta|\m S)  = \Pr( \m Y \in  B(\m S) | \theta) $, where 
\begin{align*}
B( \m S)  & = \{  \m Y : s_{i,j} >0 \Rightarrow y_{i,j}>0 , s_{i,j} = 0 \Rightarrow y_{i,j}\leq 0 \} . 
\end{align*}
To account for the censoring, we note that 
we should only infer that 
person $i$ does not positively rate person $j$
($y_{i,j}\leq 0$) if
they do not rank them
($s_{i,j}=0$)  \emph{and} 
person $i$ has unfilled nominations $(d_i< m)$. 
Our modified set of allowable $\m Y$-values can then be described by the 
conditions
\begin{align}
 s_{i,j}  >0 & \Rightarrow y_{i,j}>0   \setcounter{equation}{1} \\   
  s_{i,j} = 0 \  \mbox{and} \ d_{i}< m   & \Rightarrow y_{i,j}\leq 0   \setcounter{equation}{3}  \\
% \min_{j: s_{i,j}>0 } \{ y_{i,j}  \} & \geq  
%  \max_{j:s_{i,j}=0 } \{ y_{i,j} \} .  
\min \{ y_{i,j} : s_{i,j} > 0 \} & \geq  \max \{ y_{i,j} : s_{i,j} = 0\}
 \setcounter{equation}{10}. 
\end{align} 
The restrictions (2) and (4) are two of the three restrictions 
used to form the FRN likelihood. Restriction (11) is similar 
to restriction (3) of the FRN likelihood, in 
that it recognizes a preference ordering between ranked and 
unranked individuals, but unlike the FRN likelihood it does 
not recognize differences among ranked individuals. 

Letting $C(\m S) $ be the set of $\m Y$-values consistent with 
(2), (4) and (11), we refer to 
\[  L_C(\theta: \m S) = \Pr( \m Y \in C(\m S) |\theta) \]
as the censored binomial likelihood. 
As the censored binomial likelihood recognizes the censoring in FRN data, 
we expect it
to provide parameter estimates that do not have the biases
of the binomial likelihood estimators. On the other hand, $L_C$
ignores the information in the ranks of the scored individuals, 
and so we might expect it to provide less precise estimates 
than the FRN likelihood. To investigate these possibilities, 
we obtained $L_C$-based estimates for each of the 16 simulated 
datasets described above. The posterior mean estimates and standard 
deviations of $\v\beta$ were very similar to those obtained 
from the FRN likelihood, indicating that the censored binomial 
likelihood properly accounts for censoring in the FRN data, and 
that the information about $\v\beta$ contained in the scores of the 
ranked individuals is minimal, at least for these values of the simulation 
parameters.  To investigate this latter claim further, we performed an 
additional simulation study in which the maximum number $m$ 
of ranked individuals 
varied from 5 to 50 (out of a population of 100 individuals). 
Intuitively, the amount of information in the ranks should increase as 
the number of ranked individuals increases, and so we might expect 
the posterior distributions based on the 
FRN likelihood to be more concentrated around the true values 
than those based on $L_C$ for large values of $m$. 

\begin{figure}
\centerline{\includegraphics[width=4.5in]{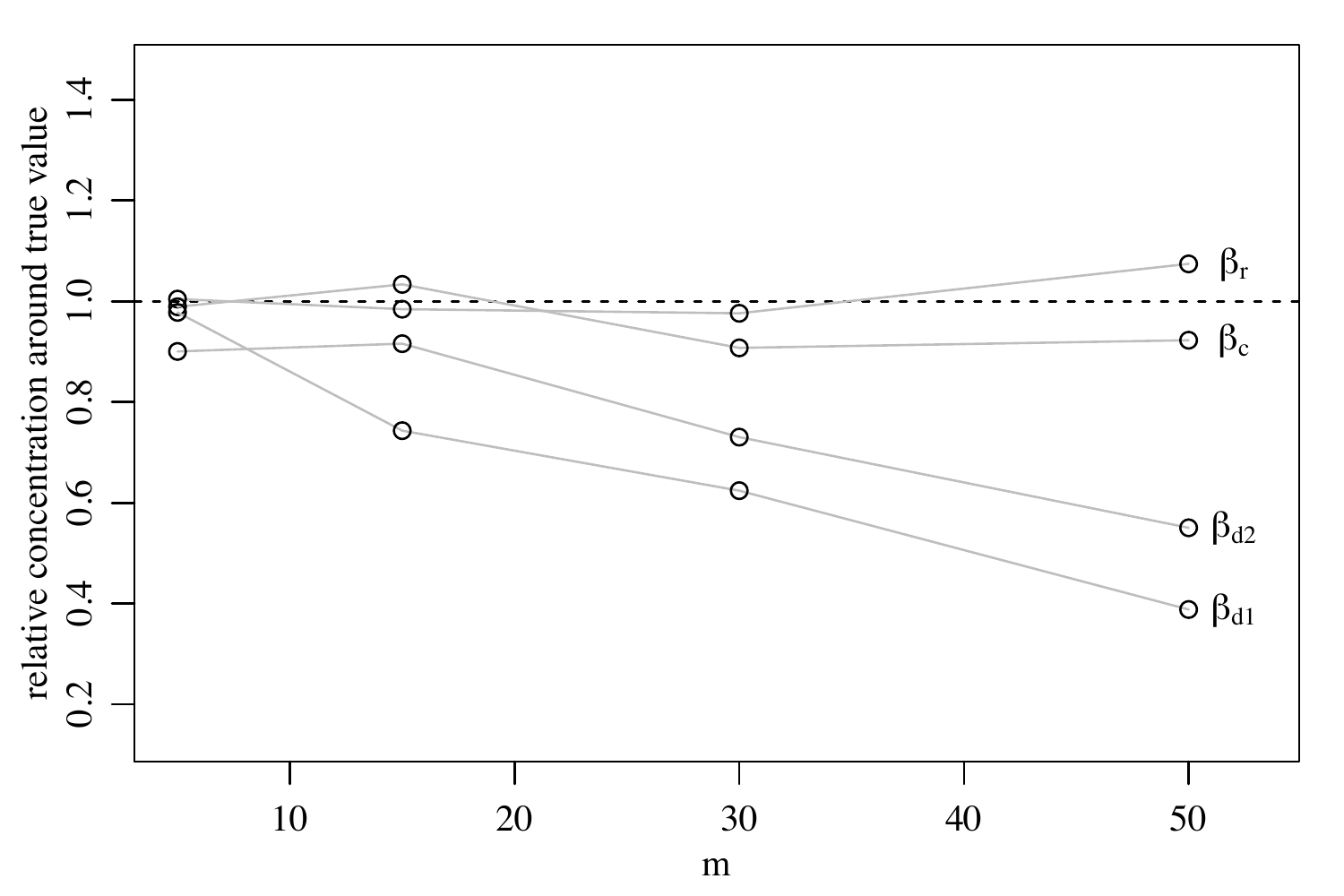}}
\caption{Posterior concentration around true parameter values.
The average of $\Exp{ (\beta -  \beta^* )^2 | F(\m S) } /
                \Exp{ (\beta - \beta^* )^2 | C(\m S) }   $  
across eight simulated datasets for each $m\in \{ 5,15,30,50\}$.  }
\label{fig:mvar}
\end{figure}

%Figure \ref{fig:mvar} show results from this study for values of 
%of $m\in \{ 5, 15, 30 , 50\}$. 
For each $m \in \{ 5, 15, 30 , 50\}$, eight datasets were simulated
as in the previous simulation study, 
using an intercept parameter  $\beta_0$  so that the average 
of the uncensored outdegrees was $m$. 
Variability in the regressors and the random effects implies that 
 for each $m$, some 
simulated individuals had uncensored 
outdegrees $\tilde d_i$ above the censored value 
$m$ and some had outdegrees below. 
For each regression parameter $\beta$ in the model and each simulated dataset, 
we computed  the ratio
$\Exp{ (\beta - \beta^*)^2  | F(\m S) } / \Exp{ (\beta - \beta^*)^2  | C(\m S) }$,
where $\beta^*$ is the true value of the parameter  (here, $\beta^* = 1$
for each parameter except the intercept).  This ratio 
measures the relative concentrations of the posterior distributions
$p(\beta | \m Y \in F(\m S) $ and $p(\beta |\m Y \in C(\m S))$ around the 
 true value of the parameter. 
These ratios were averaged across the eight simulated datasets for 
each value of $m$, and plotted in Figure \ref{fig:mvar}.  
The plots indicate that, for the parameter values 
considered here, the censored binomial likelihood 
suffers no noticeable information loss in terms 
of estimating the row and column regression parameters 
$\beta_r$ and $\beta_c$, but 
provides substantially less precise parameter estimates for 
the dyadic-level parameters $\beta_{d1}$ and $\beta_{d2}$ at 
high values of $m$. 
However, we note that this loss in precision does not 
appear to be appreciable 
until $m$ is a quarter to a third of the number $n$ of individuals in 
the network.  These results suggest that  
for  FRN surveys where $m$ is substantially smaller 
than $n$, the 
majority of the information about the regression parameters comes 
from distinguishing between between ranked and unranked individuals, 
and that the relative ordering among the ranked individuals 
provides at most a modest amount of additional information. 
For such surveys, 
the censored binomial likelihood may provide an adequate 
approximation to inferences that would be obtained under 
the FRN likelihood.

\section{Analysis of AddHealth data}
As described in the Introduction, one component of the AddHealth 
study included fixed rank nomination surveys administered to a 
national sample of high schools.  Within each school, each participating 
student was asked to  nominate and rank up to five same-sex friends
and five friends of the opposite sex. Students were also asked to 
provide information about a variety of their own characteristics, 
such as ethnicity, academic performance, smoking and drinking behavior
and extra-curricular activities.   
To describe the relationships between an individual's  characteristics 
and the friendship nominations they send and receive, we 
fit the social relations regression model  
(\ref{eq:srmab}), with a mean model given by
\begin{align*}
\Exp{ y_{i,j} | \v \beta, \v x_{i,j} }   & =  \v \beta^T \v x_{i,j} 
 =   {\v \beta}_r^T \v x_{r,i } + {\v \beta}_c^T \v x_{c,j }  +
          {\v \beta}_d^T \v x_{d,i,j } , 
\end{align*}
where $\v\beta_r$, $\v\beta_c$ and  $\v\beta_d$ are vectors of unknown regression 
coefficients, corresponding to 
row-specific, column-specific and dyad-specific regressors. 
We fit such a model to 
both the male-male and female-female FRN networks  of 
7 schools from the AddHealth study,
where the schools were chosen based on their high within-school survey participation 
rates. 
Based on an initial exploratory data analysis of these 14 FRN networks, 
the  following row, column and dyadic regressors were selected:
%following regressors were chosen to be included in the model:
\begin{align*}
\v x_i  & =  ( {\tt rsmoke}_{i}, {\tt rdrink}_{i}, {\tt rgpa}_{i}   ) \\
\v x_j  &=   ( {\tt csmoke}_{j}, {\tt cdrink}_{j}, {\tt cgpa}_{j} )  \\
\v x_{i,j} &=  ( {\tt dsmoke}_{i,j}, {\tt ddrink}_{i,j}, {\tt dgpa}_{i,j}  ,
{\tt dacad}_{i,j},{\tt darts}_{i,j}, {\tt dsports}_{i,j} , {\tt dcivic}_{i,j},
   {\tt dgrade}_{i,j},{\tt drace}_{i,j} ) 
\end{align*}
A description of the variables is as follows:
\begin{description}
\item[behavioral characteristics:]  
Self-reported GPA and 
smoking and drinking activity were ranked among all students 
of a given sex within a school and converted to 
normal $z$-scores via a quantile transformation. 
These $z$-scores were included both as 
 row-specific regressors ({\tt rsmoke, rdrink, rgpa}) and 
column-specific regressors ({\tt csmoke, cdrink, cgpa}),
and formed the basis of dyadic interaction  terms
  ({\tt dsmoke, ddrink, dgpa}). For example, 
{\tt dsmoke}$_{i,j}$ = {\tt rsmoke}$_{i} \times$ {\tt csmoke}$_{j}$. 
\item[extracurricular activities:]
Participation in school-sponsored extracurricular activities  
was categorized by activity type (academic, artistic, sports, civic). 
%and the number of activities in each category was included as a
%column specific regressor 
%({\tt cacad}, {\tt carts},{\tt csports},{\tt ccivic}).  
The numbers of activities of each type jointly participated in by 
pairs of students were included as dyadic regressors 
({\tt dacad}, {\tt darts}, {\tt dsports}, {\tt dcivic}).  
For example, {\tt dsports}$_{i,j}$ is the number of sports 
in which both student $i$ and student $j$ participated. 
\item[demographic characteristics:] For each pair of students $(i,j)$, 
a binary indicator of same grade ({\tt dgrade}) and 
a Jaccard measure of racial similarity 
({\tt drace})
were included as dyadic regressors. 
\end{description}

%\begin{align*}
%+ a_i + b_ j +\epsilon_{i,j} \\
%\left( ( \begin{smallmatrix} \epsilon_{i,j}  \\ \epsilon_{j,i} \end{smallmatrix} ) , i\neq j \right) & \sim 
% \mbox{ i.i.d.\ normal}(\v 0,\sigma^2 (\begin{smallmatrix} 1 & \rho \\ \rho & 1 \end{smallmatrix} )  ) 
%\end{align*}

As the data were obtained using an FRN study design, it seems most 
appropriate to estimate the regression coefficients 
$\v \beta=(\v \beta_r,\v \beta_c, \v\beta_d)$ using the FRN likelihood 
described in Section 2. Also  of interest is a comparison
of such estimates to those obtained using the binomial and rank likelihoods, 
in order to see if the relationships between the estimates are similar to those 
seen in the simulation study in Section 3.2. 
To this end, we obtained parameter estimates and 
confidence intervals of $\v \beta$ 
for each of the 14 FRN networks and each  of the three likelihoods. 
In the interest of brevity, 
we give details on the data and results for 
the male-male and female-female network for only one school, 
and briefly summarize the results for the remaining 12.

\begin{figure}
\centerline{\includegraphics[width=6.5in]{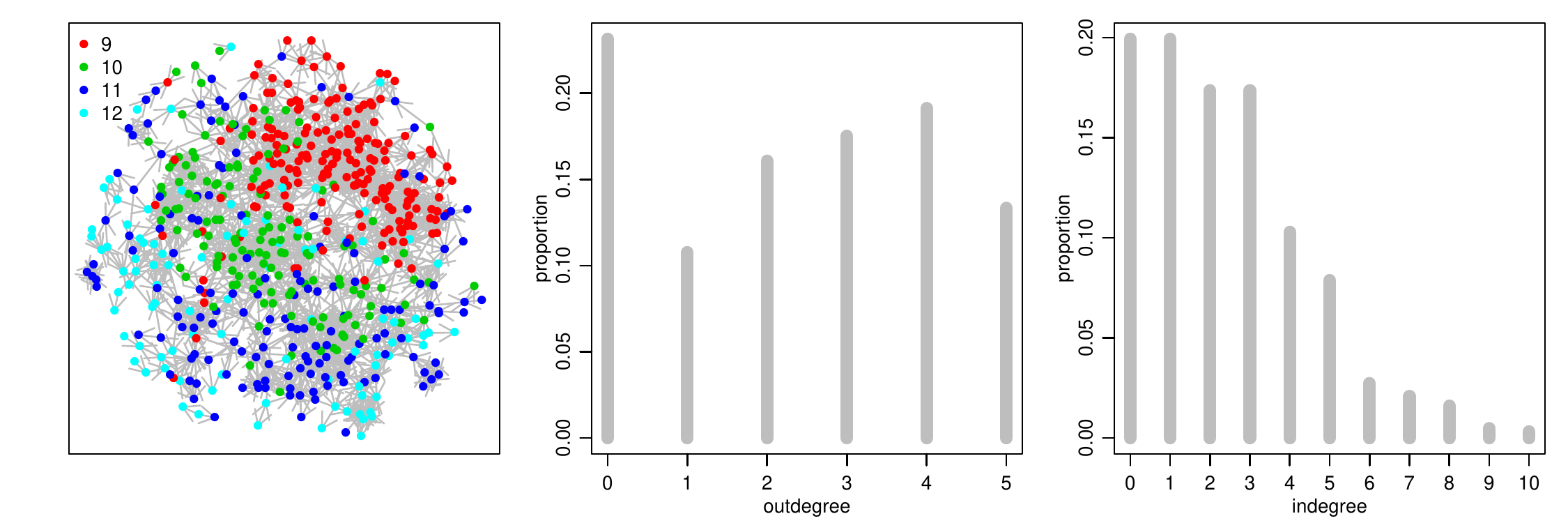}}
\caption{Male nomination network. }
\label{fig:mnom}
\end{figure}

\begin{figure}
\centerline{\includegraphics[width=6.5in]{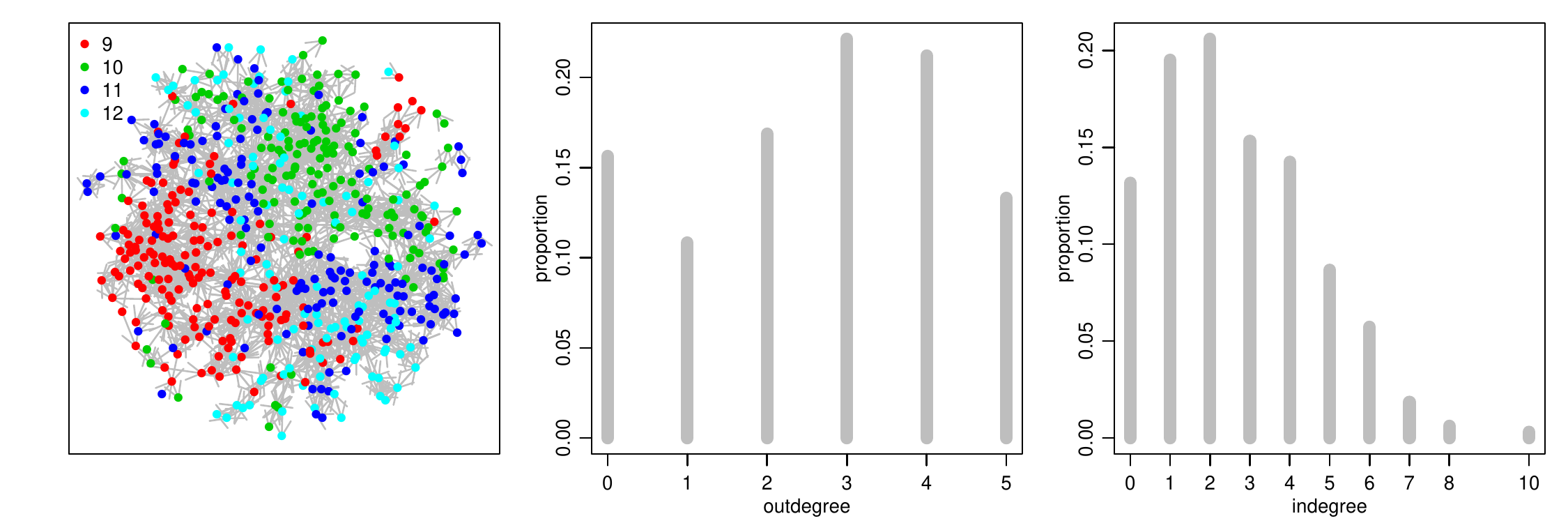}}
\caption{Female nomination network. }
\label{fig:fnom}
\end{figure}

Graphical descriptions of the male-male and female-female FRN  
networks of the largest of the 7 schools 
are presented in Figures \ref{fig:mnom}
 and \ref{fig:fnom} respectively. The networks are based on data from 
622 male and 646 female study participants. The first plot in each 
row consists of a graph with edges representing the friendship nominations
and nodes representing the students, color-coded by grade. 
The second and third plots give the degree distributions, i.e.\ 
the empirical distributions of the 
number of nominations made to other survey participants (outdegree) 
and number of nominations received by other survey participants
(indegree).   All outdegrees are less than or equal to  
5, reflecting the fact that each student was allowed to make at most 
5 nominations. A substantial number of students also report 
0 friendships to other survey participants, but this should not 
be taken to mean that they have zero friendships: 
A substantial fraction of the
friendship nominations of survey participants were to
students in the school who did not participate in the survey 
(22\% for this school), 
or to individuals outside the school entirely. 
As no information is available for these out-of-survey individuals, 
we cannot include them in the model directly. However, the FRN 
likelihood can be modified to accommodate this information 
indirectly, by 
%conditioning on the number of 
recognizing that the number of out-of-survey friendship 
ties alters how within-survey ties are censored:
%out-of-survey friendship ties: 
If individual $i$ 
makes $d_i^o$ out-of-sample nominations, they have $m_i = m- d^o_i $ 
remaining nominations to allocate to within-survey friendships. 
If individual $i$ makes $d_i^o$ out-of-survey nominations 
and $d_i < m-d^o_i$ within-survey nominations,  then they 
are indicating that 
they do not have any further positive within-survey relationships. 
In contrast, if $d_i = m-d^o_i$ then 
this individual's relationships are censored 
and we do not have information 
on the presence or absence of additional within-survey 
relationships. Accounting for this censoring information 
can be made by modifying Equation 4 defining the FRN likelihood  to 
be 
\[ s_{i,j} = 0 \ \mbox{and} \ d_i < m_i \Rightarrow y_{i,j} \leq 0, \]
where the only change from Equation 4 is that the 
maximum (within-survey) outdegree is now the individual-specific 
value $m_i$ as opposed to being a common value $m$.

\begin{figure}
\centerline{\includegraphics[width=6.5in]{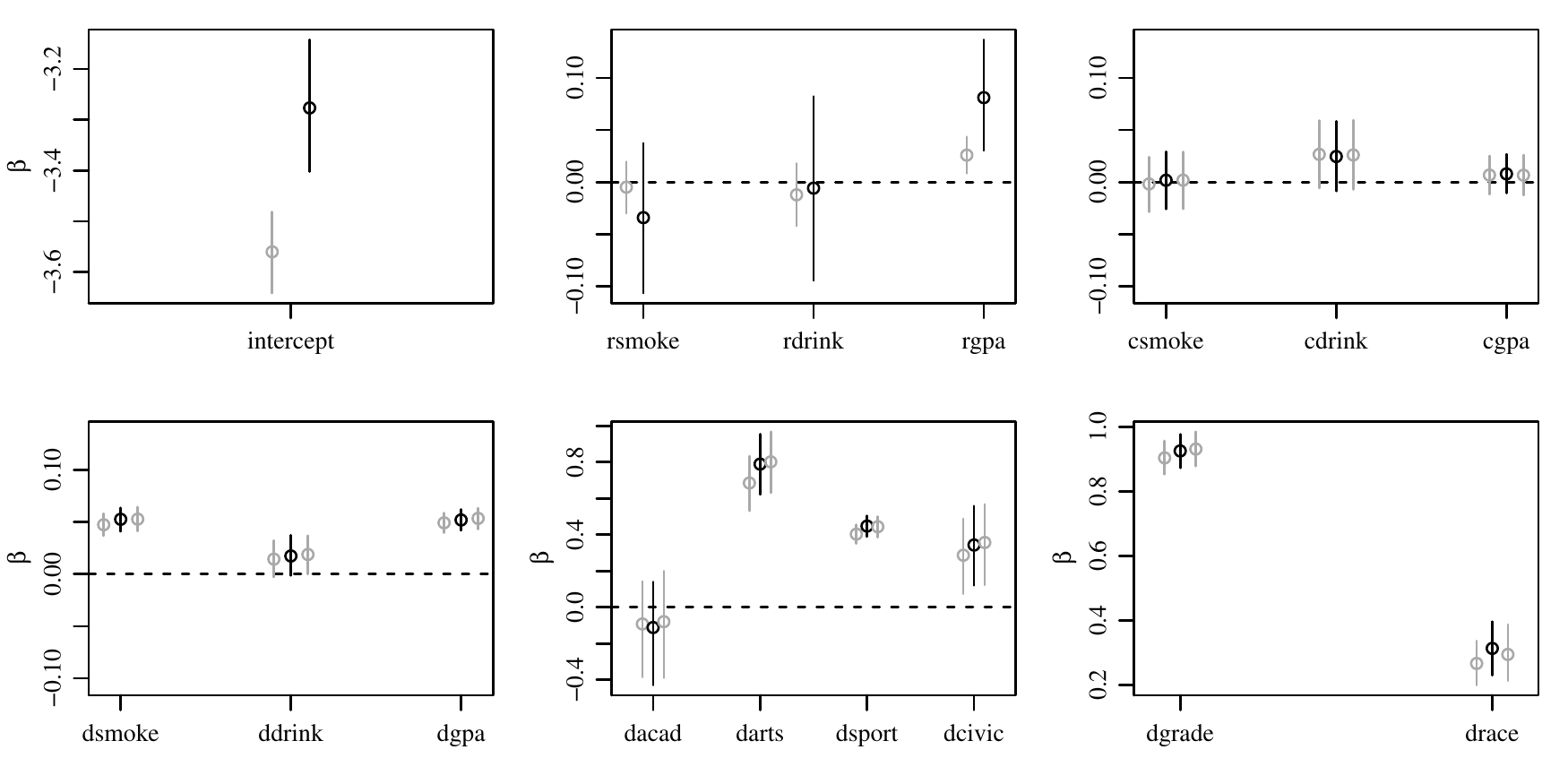}}
\caption{Parameter estimates and confidence intervals for $\v \beta$
in the male-male network. Each group of intervals represents
from left to right the intervals obtained from the binomial, FRN and
rank likelihoods, respectively. The rank likelihood
does not provide parameter estimates for an intercept or for the row-specific
effects {\tt rsmoke}, {\tt rdrink}, {\tt rgpa}. }
\label{fig:mmcoef}
\end{figure}

\begin{figure}
\centerline{\includegraphics[width=6.5in]{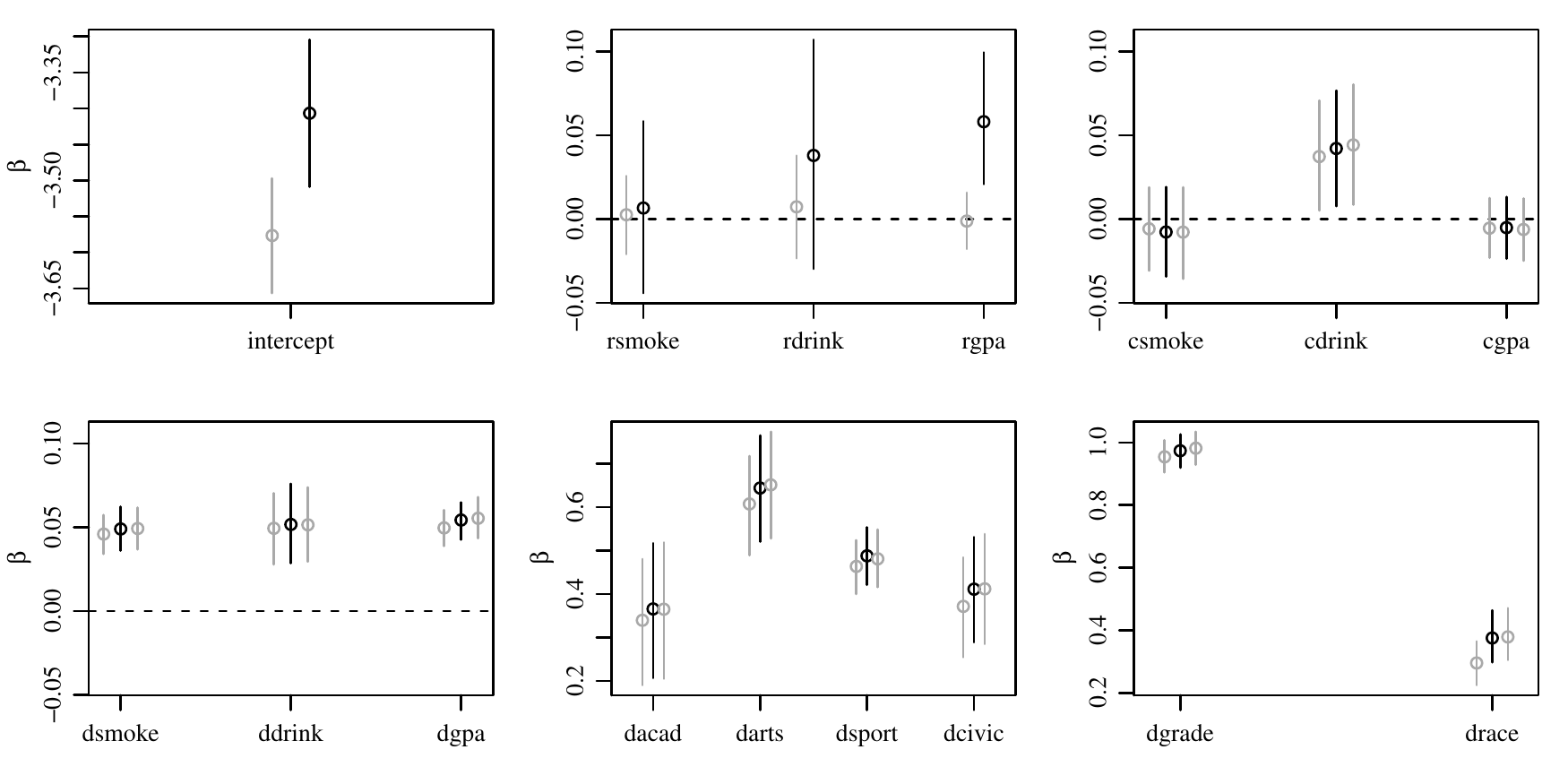}}
\caption{Parameter estimates and confidence intervals for $\v \beta$
in the female-female network.   }
\label{fig:ffcoef}
\end{figure}

Using the MCMC algorithms described in Section 2.2, 
we obtained parameter estimates and confidence intervals of 
$\v\beta$ for both the male-male and female-female networks, 
using the FRN, binomial and rank likelihoods. 
The Markov chains appeared to converge very quickly. 
After an initial burn-in period of 500 iterations, 
each Markov chain was run for an additional 500,000 iterations, 
from which 
every 25th iteration was saved, resulting in 20,000 simulated 
values for each parameter with which to make inference. 
Average effective sample sizes 
%(the equivalent number of independent iterations) 
across parameters and Markov chains 
were 5,288, 5,764 and 2,079
for the FRN,  binomial and rank likelihoods, respectively. 
Posterior medians and 95\% confidence intervals for all regression 
parameters are shown in Figures \ref{fig:mmcoef} and 
   \ref{fig:ffcoef}. 
Point estimates and confidence intervals based on the FRN likelihood 
suggest that for both males and females, an individual's GPA ({\tt rgpa}) is 
positively associated with their evaluation of other individuals 
as friends, and that increased drinking behavior ({\tt cdrink}) seems to 
be positively associated with an individual's popularity. 
Additionally,
the dyadic effect estimates indicate that, on average, similarity between 
two individuals by just about any measure increases their evaluation 
of each other.

Parameter estimates and confidence intervals based on the 
binomial and rank likelihoods generally provide similar 
conclusions about the effects,  the exception being the 
intercept and  row effects. 
Intercept estimates under the binomial likelihood are lower than 
those under the FRN likelihood, as they fail to recognize the censoring in 
outdegree. 
For both males and females, coefficient estimates
for the row effects are generally closer to zero  under the binomial 
likelihood than the FRN likelihood, and the confidence intervals 
are substantially narrower.  
In particular, FRN likelihood
 confidence intervals for {\tt rdrink} and {\tt rgpa}  
in the female network 
(the second plot of Figure  \ref{fig:ffcoef})  are centered around 
positive values, whereas the corresponding binomial likelihood intervals
are essentially centered around zero. 
These phenomena 
are similar to  the 
patterns of bias seen in 
the simulation
study in Section 3.2, and predicted by the analytical approximation
in Section 3.1.

We fit the same model to the male-male and female-female network 
of six additional schools (12 additional networks). Generally speaking, 
the 
same pattern 
of differences between the  different estimators 
appeared for these schools as for the school analyzed above and 
in the simulation study:  As compared to the FRN likelihood, 
the binomial likelihood 
estimated the intercept as being too low and  the row effects as too close to 
zero with overly-narrow confidence intervals.  Additionally, 
parameter estimates  for 
dyadic effects  
that were not mean-centered (such as {\tt dgrade} and {\tt drace})  were also too close to zero. 
These results are summarized in Table \ref{tab:relman}. 
For each effect type, we computed the (geometric) average 
ratio of the magnitude of the parameter estimate under the FRN likelihood 
to those under the binomial and rank likelihoods.  
Besides having larger (negative) intercept estimates than the FRN likelihood, 
the binomial likelihood generally had estimates with smaller magnitudes, 
especially for the row effects and the non-mean-zero dyadic 
effects {\tt dgrade} and {\tt drace}. 
We also computed the  average
ratio of the confidence interval widths under the different likelihoods. 
Interval widths were generally 
similar across likelihoods, the main exception being that the 
interval widths 
for the row effects 
under the binomial likelihood were on average three 
times narrower than the intervals obtained from the FRN likelihood, 
similar to what was seen in the simulation study. 
In contrast, the rank likelihood provides parameter estimates and 
interval widths that are comparatively close to those from the FRN likelihood. 

\begin{table}[ht]
\begin{center}
\begin{tabular}{r |ccccc} 
     &  \multicolumn{5}{c}{Effect type} \\
Likelihood   &  intercept   &  row  &  column    &
  mean-zero dyadic  &   other dyadic  \\  \hline
binomial & 0.89, 1.68  & 2.22, 2.95 &  1.02, 1.03 &  1.06, 1.06 &  1.20, 1.09   \\
rank     &  NA,NA  & NA,NA          & 1.05, 0.98  & 0.99, 0.99 & 1.06, 0.98
\end{tabular} 
\end{center}
\caption{Average  relative magnitudes of parameter estimates (first number) and
confidence interval widths (second number) 
from the FRN likelihood as compared to
the binomial and rank likelihoods. The rank likelihood does not estimate
an intercept or row effects.  }
\label{tab:relman}
\end{table}

\section{Discussion}
A popular way to represent relational data is 
as a graph, i.e.\ a list of edges between a set of nodes. 
Such representations often entail dichotomizations of non-binary, ordinal relational data, and a loss of the context in which the data were gathered. 
Statistical methods based solely on the graphical representation of the 
data run the risk of being inefficient and misleading. 
In this article, we have shown how a binary likelihood
that uses only the graphical representation of a fixed rank 
nomination (FRN) dataset can provide incorrect inferences for a
variety of model parameters. Specifically, in a social 
relations regression model, the 
binary likelihood can substantially underestimate
the effects
 of regressors with variation among the nominators
of relations.  This includes  characteristics of the
nominators of ties, as well as dyadic indicators of group co-membership
between the nominators and nominees.

Such problems can be avoided by use of  a likelihood function based 
on the data collection scheme. In this article, we have developed 
a likelihood that accounts for the censored and 
ordinal nature of  FRN data. In a simulation study, 
parameter estimates based on this 
FRN likelihood were shown to lack the biases present in  
estimates based on the binary likelihood. Additionally, 
the FRN likelihood was seen to provide more precise inference 
for the coefficients of dyadic-level regressors when the 
number of possible nominations was large. However, 
a modified  binary likelihood  that accounted for the 
data censoring was seen to provide inference that was 
roughly as accurate as that provided by the FRN likelihood when the 
maximum number of nominations was small compared to the total 
number of individuals in the network. 
This result suggests that there may not be much information to be gained in
FRN surveys by  asking survey respondents to rank their nominations.

Our analytical and empirical comparisons were based on 
the social relations regression model, a model that does not explicitly 
represent network features such as transitivity, clustering or 
stochastic equivalence. A popular class of statistical models 
that can capture a wider variety of patterns in 
uncensored, binary network relations are exponentially 
parameterized random graph models (ERGMS) \citep{frank_strauss_1986,
wasserman_pattison_1996}. In theory, ERGM models 
could be used to model censored relations, for example, by treating the 
observed data as a censored version of network with unrestricted 
outdegrees generated from an ERGM model. However, 
estimating parameters in such a framework could be computationally 
prohibitive. Additionally, ERGMS are explicitly graph models  
for binary data, and do not accommodate valued, ranked relations. 
However, recent work by \citet{krivitsky_butts_2012} has extended 
the ideas behind ERGMs to a class of exponential family models 
for ranked relational data. 

An alternative to ERGM models are latent variable models that treat 
the data from each dyad as conditionally independent given some 
unobserved node-specific latent variables. Such models extend the 
SRM  given in (10) as 
\[ 
y_{i,j} = \v\beta^T \v x_{i,j} + a_i + b_j + f(u_i,v_j) + 
 \epsilon_{i,j}, 
\]
where $f$ is a known function and $(u_1,\ldots, u_n)$ and 
$(v_1,\ldots, v_n)$ are sender- and receiver-specific 
latent variables. 
A version of the stochastic blockmodel 
\citep{nowicki_snijders_2001}
follows when the latent variables take on a fixed number of 
categorical values. 
 Alternatively, taking $f(u_i, v_j) = 
  u_i^T v_j$, with $u_i$ and $v_j$ being low-dimensional vectors, 
gives a type of latent factor model
\citep{hoff_2005a,hoff_2009c}. 
Models such as these can capture 
patterns of stochastic equivalence, transitivity and 
clustering often found in relational datasets. 
Parameter estimation for such models using the FRN likelihood 
can be achieved within the MCMC framework described in 
Section 2.2 with the addition of steps for updating 
values of the  latent variables.

The simulation study and network analyses in this article were 
implemented in the open source 
{\sf R} statistical computing environment using the 
{\tt amen} package, available at \url{http://cran.r-project.org/web/packages/amen/}.  Replication code for the simulation study is available at the first 
author's website, \url{http://www.stat.washington.edu/~hoff/}.

\bibliographystyle{plainnat}
%\bibliography{/Users/hoff/Dropbox/SharedFiles/refs}
\bibliography{refs}

\end{document}